\documentclass[10pt,conference]{IEEEtran}
\IEEEoverridecommandlockouts
\usepackage[utf8]{inputenc}

\usepackage[
    backend=biber,
    sorting=none,
    style=ieee
]{biblatex}

\usepackage{graphicx}
\graphicspath{{hcw/figures}}

\usepackage[normalem]{ulem}
\usepackage{subcaption}
\usepackage{listings}
\usepackage[hidelinks]{hyperref}
\usepackage{multirow}
\usepackage{xcolor}

\begin{document}

\title{CEDR-API: Productive, Performant Programming of Domain-Specific Embedded Systems\\
\thanks{This material is based on research sponsored by Air Force Research Laboratory (AFRL) and Defense Advanced Research Projects Agency (DARPA) under agreement number FA8650-18-2-7860. The U.S. Government is authorized to reproduce and distribute reprints for Governmental purposes notwithstanding any copyright notation thereon. The views and conclusion contained herein are those of the authors and should not be interpreted as necessarily representing the official policies or endorsements, either expressed or implied, of Air Force Research Laboratory (AFRL) and Defence Advanced Research Projects Agency (DARPA) or the U.S. Government. 
}
}

\makeatletter
\newcommand{\linebreakand}{%
  \end{@IEEEauthorhalign}
  \hfill\mbox{}\par
  \mbox{}\hfill\begin{@IEEEauthorhalign}
}
\makeatother

\author{
\IEEEauthorblockN{
    Joshua Mack
}
\IEEEauthorblockA{
    \textit{ECE, The University of Arizona} \\
    Tucson, Arizona, USA \\
    jmack2545@arizona.edu
}
\and
\IEEEauthorblockN{
    Serhan Gener
}
\IEEEauthorblockA{
    \textit{ECE, The University of Arizona} \\
    Tucson, Arizona, USA \\
    gener@arizona.edu
}
\and
\IEEEauthorblockN{
    Sahil Hassan
}
\IEEEauthorblockA{
    \textit{ECE, The University of Arizona} \\
    Tucson, Arizona, USA \\
    sahilhassan@arizona.edu
}
\linebreakand
\IEEEauthorblockN{
    H. Umut Suluhan
}
\IEEEauthorblockA{
    \textit{ECE, The University of Arizona} \\
    Tucson, Arizona, USA \\
    suluhan@arizona.edu
}
\and
\IEEEauthorblockN{
    Ali Akoglu
}
\IEEEauthorblockA{
    \textit{ECE, The University of Arizona} \\
    Tucson, Arizona, USA \\
    akoglu@arizona.edu
}
}
\maketitle

\begin{abstract}
As the computing landscape evolves, system designers continue to explore design methodologies that leverage increased levels of heterogeneity to push performance within limited size, weight, power, and cost budgets.
One such methodology is to build Domain-Specific System on Chips (DSSoCs) that promise increased productivity through narrowed scope of their target application domain.
In previous works, we have proposed CEDR, an open source, unified compilation and runtime framework for DSSoC architectures that allows applications, scheduling heuristics, and accelerators to be co-designed in a cohesive manner that maximizes system performance.
In this work, we present changes to the application development workflow that enable a more productive and expressive API-based programming methodology.
These changes allow for more rapid integration of new applications without sacrificing application performance. 
Towards the design of heterogeneous SoCs with rich set of accelerators, in this study we experimentally study the impact of increase in workload complexity and growth in the pool of compute resources on execution time of dynamically arriving workloads composed of real-life applications executed over architectures emulated on Xilinx ZCU102 MPSoC and Nvidia Jetson AGX Xavier. 
We expand CEDR into the application domain of autonomous vehicles, and we find that API-based CEDR achieves a runtime overhead reduction of 19.5\% with respect to the original CEDR.

\end{abstract}


\section{Introduction} \label{sec:intro}

As technology scaling reaches its limits, system designers are exploring an increasingly diverse range of methodologies for building systems that can maximize their compute performance within limited size, weight, power, and cost budgets.
One such methodology in the literature is the design and fabrication of Domain-Specific System on Chip (DSSoC) devices.
The motivation of such devices is fairly simple: in a general-purpose computing context, heterogeneous computing systems are difficult to program and utilize effectively.
The hypothesis with DSSoC devices, then, is that by restricting the applications used on a given system to a particular domain, it becomes more feasible to build productive software and programming abstractions for the finalized hardware.

We believe that such software and programming abstractions have a number of key requirements.
These abstractions should support scenarios where multiple users can coexist and interleave their applications across the DSSoC's heterogeneous pool of processing elements (PEs) while utilizing compute resources effectively in a dynamic way.
Many of the predominant heterogeneous compute frameworks such as CUDA~\cite{nickolls_cuda_2008}, OpenCL~\cite{stone_opencl_2010}, or SYCL~\cite{reyes_2020_sycl} assume an environment where an application expert performs offline analysis across a number of possible implementations for a given application, determines the optimal static mapping for all computational kernels in that application, and produces a fixed binary that represents a single, expertly-tuned instance of the application.
In an environment of widespread heterogeneous computation, these greedy, inflexible mappings ignore runtime resource contention and will clearly lead to system inefficiencies when such programs are required to share system resources with an arbitrary number of other heterogeneous applications. 
One might expect that resource contention issues would be solved by the operating system, but Roscoe argues~\cite{roscoe_2021_rediscover} that the rate of architectural advancement for modern SoCs has tremendously outpaced the work in operating systems to meet them.
Hence, to handle this, we believe that DSSoC software abstractions must be coupled with intelligent, intermediate runtime systems that are capable of arbitrating or scheduling requests from all applications to the PEs across the DSSoC.
To enable this runtime to effectively arbitrate resources among applications, each application must also be capable of mapping its computational kernels to as wide of a number of the system's heterogeneous processing elements as possible.
Namely, the programming abstractions provided to the user should be agnostic from the underlying hardware as this provides a number of key benefits: it allows for maximum flexibility when the runtime is performing its scheduling, and it allows the user to program for the DSSoC in a productive manner without needing to become a hardware expert in the process.

In prior work, we have explored the development of such a runtime through a framework called CEDR, a Compiler-Integrated, Extensible, DSSoC Runtime~\cite{mack_userspaceemulation_2020-nonblind,mack_cedrtecs_2023-nonblind}.
CEDR has been designed to holistically target the joint challenges of choosing the composition and capabilities of accelerators on a DSSoC, determining the optimal scheduling heuristic for an application domain, and providing a productive set of software abstractions for programmers writing applications for DSSoCs.
The primary contribution of CEDR thus far is its status as an open source ecosystem~\footnote{Available at: \url{https://github.com/UA-RCL/CEDR}} that integrates compile-time application analysis with a Linux userspace-based runtime system that addresses the aforementioned challenges.
The programming model for CEDR relies on a Directed Acyclic Graph (DAG)-based program representation where each node in the graph represents schedulable unit of computation, and edges represent temporal dependencies.
While this representation has proven itself to be a sufficiently powerful format to meet the aforementioned challenges with clear support for notions of concurrency and heterogeneity, it does have a number of limitations with regards to programmer productivity.
First, despite a large amount of progress in compilation tooling, it is still comparatively difficult for many programmers to express applications in performant, DAG-based representations.
Additionally, a DAG-based application format cannot accurately capture the control flow structures of many programs in a way that preserves scheduler flexibility without a mechanism to enable conditional DAG dependencies.
In this work, we seek to address these limitations through the integration of a simplified API-based programming model that better aligns with traditional practices in programming for homogeneous computing systems.
By incorporating both blocking and non-blocking APIs, we are able to make this change without impacting the ability to write performant applications and thus the utility that CEDR provides for hardware designers or scheduling heuristic developers seeking to stress their systems.
The primary contributions of this work are as follows:

\begin{itemize}
    \item We modify the programming model of CEDR to allow for a more productive API-based application workflow. 
    \item We integrate non-blocking APIs that allow for performance programmers to maximally exploit opportunities for parallelism in their code.
    \item We expand CEDR to the domain of autonomous vehicles and incorporate lane detection.
    \item We perform hardware experiments with increasing heterogeneity and expose scenarios in which parallelism of heterogeneous accelerators is limited by the availability of homogeneous CPU cores.
\end{itemize}

The rest of the paper is organized as follows: in Section~\ref{sec:cedr-api}, we explore the details of the modified API-based CEDR. 
In Section~\ref{sec:experimental-setup}, we detail the workload and system configurations that will be leveraged in our experimental analysis. 
In Section~\ref{sec:experiments}, we analyze our modified CEDR runtime across a variety of workload and system configurations. 
In Section~\ref{sec:related-work}, we include an updated analysis of CEDR's standing within the literature.
Finally, in Section~\ref{sec:conclusion}, we summarize our findings and conclude with avenues for future work.

\section{CEDR-API} \label{sec:cedr-api}

\subsection{Background}
\label{sec:background}
\begin{figure}[t]
    \centering
    \includegraphics[width=0.9\linewidth]{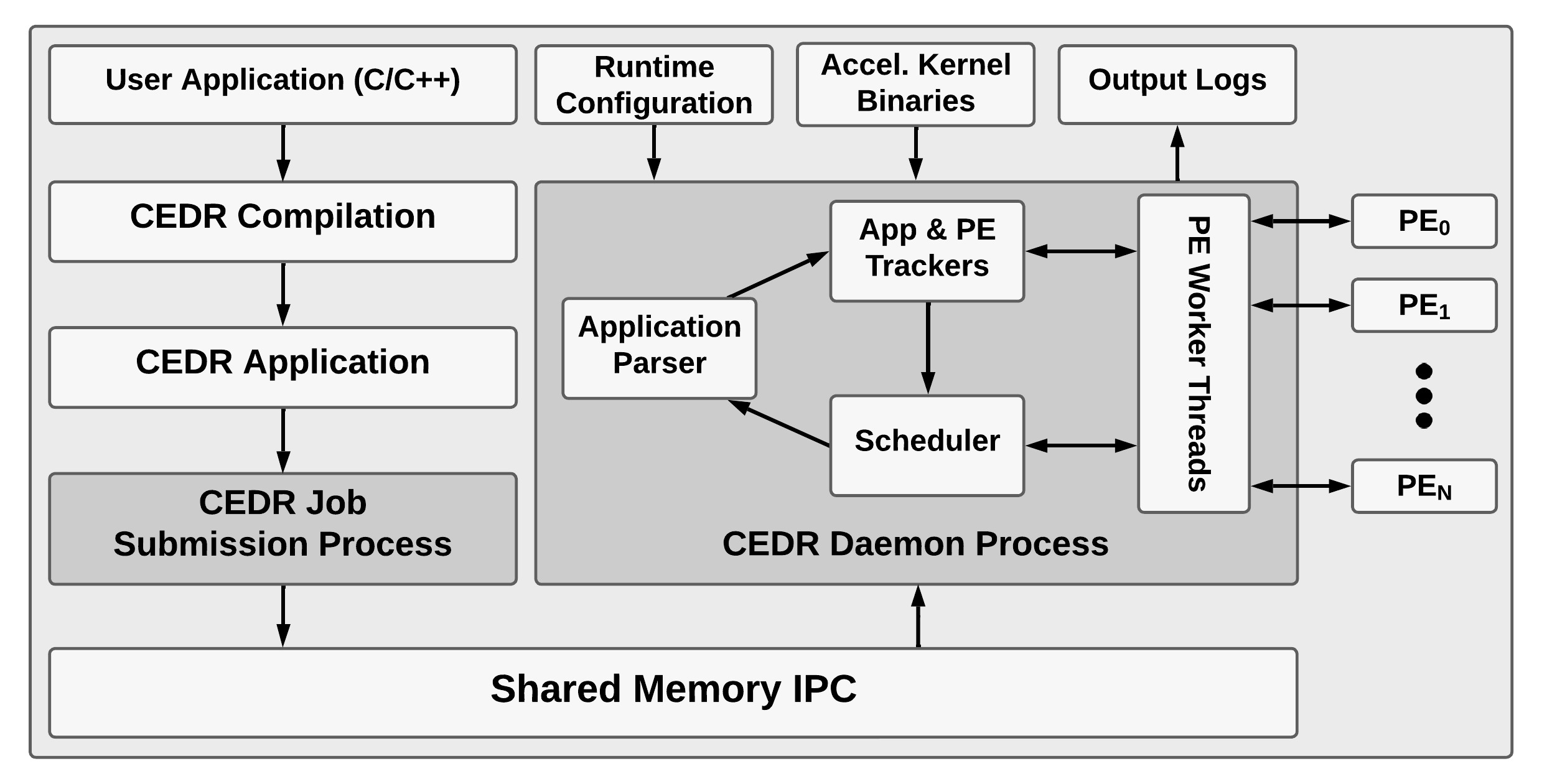}
    \caption{An overview of CEDR's integrated compiler and runtime system.}
    \label{fig:cedr_runtime}
     \vspace{-4mm}
\end{figure}

For completeness, we begin with an overview of how the existing CEDR runtime operates.
Shown in Fig.~\ref{fig:cedr_runtime}, CEDR is broadly composed of two components: a compilation workflow and a runtime workflow.
The compilation workflow is used to transform user applications into applications that CEDR can execute.
For details about this process in the baseline runtime, we refer readers to Section 2.2 of Mack et al.~\cite{mack_cedrtecs_2023-nonblind}.
At the end of this process, a CEDR Application is produced that consists of a shared object binary and a JSON-based DAG.
The shared object contains a number of functions representing the computational nodes within the DAG, but it does not contain application-level control flow information.
Instead, the JSON captures temporal dependencies between nodes and high level control flow of the user's application.
This pair of objects is submitted to the runtime, referred to as the ``CEDR Daemon Process'' in Fig.~\ref{fig:cedr_runtime} via inter-process communication (IPC).

The CEDR Runtime consists of a few key components: a number of worker threads, a ready queue of tasks, and a main event loop that receives, parses, launches, and manages applications.
Each PE in the system -- accelerator or CPU core -- is paired with a worker thread that manages executing tasks on that compute resource.
In the case of CPU cores, each worker thread is assigned via its processor affinity to run on the corresponding resource.
However, for accelerator cores, their respective worker thread is assigned via its affinity to some CPU core in the system, and that CPU core is responsible for coordinating any configuration updates or data transfers that accelerator requires.
Within the main event loop, CEDR periodically pushes work to these threads by scheduling tasks to them from its ready queue of tasks according to a user-defined heuristic as part of the program's \emph{Runtime Configuration} input.
The \emph{Runtime Configuration} also allows the user to enable or disable features such as PAPI-based performance counters~\cite{terpstra2010CollectingPerformance}.
As tasks are completed, the worker threads signal completion back to the main thread and JSON DAG dependencies of those tasks are then pushed to the back of the ready queue.
To support heterogeneous execution, when a task is scheduled to a given resource, CEDR dynamically updates that task's function pointer such that its worker thread invokes a function that is compatible with that resource.
As applications are submitted over the IPC channel, they are parsed and their executions are started by placing the head nodes of their DAGs into this same ready queue.
This process continues indefinitely until an IPC command is received that tells CEDR to shutdown at which point it serializes all the logs it has collected relating to task execution, performance counter measurements, and so on for later offline analysis by the user.

\begin{figure}[tbh]
    \centering
    \includegraphics[width=0.75\linewidth]{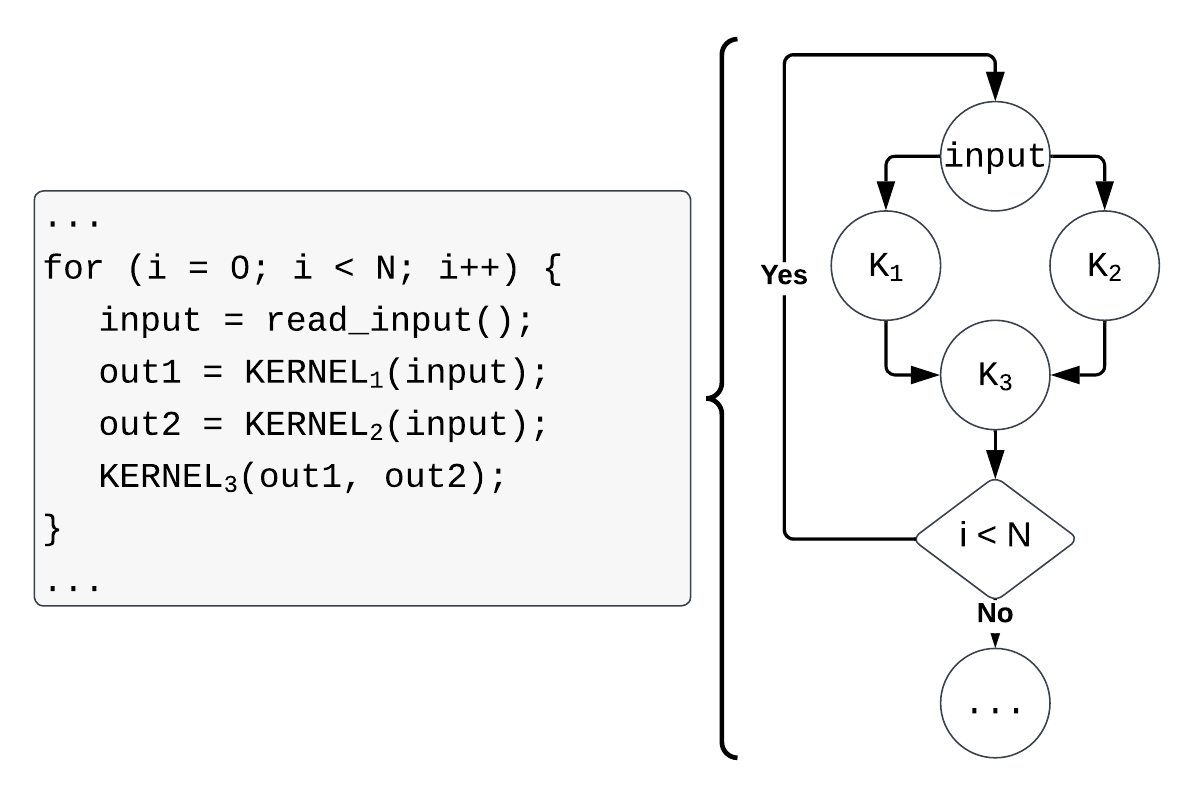}
    \caption{An example of an application structure that cannot be represented in the previous DAG-based representation without compressing it to a single node and losing the ability to schedule each kernel independently.}
    \label{fig:limitations_of_dag}
     \vspace{-4mm}
\end{figure}

\subsection{Limitations of the Existing Runtime}
\label{sec:limits}
To motivate the contributions of this work, we first begin by discussing limitations in the types of applications that can be represented with a DAG-based format in the previous release of CEDR.
Due to the acyclic nature of DAG-based applications, this format is unable to represent control flow concepts of iteration or conditional branching.
This leads to issues when trying to schedule applications that have structures like that shown in the left half of Fig.~\ref{fig:limitations_of_dag}.
Kernel\textsubscript{1}, Kernel\textsubscript{2}, and Kernel\textsubscript{3} may all be individually compatible with accelerators on the system, but because a DAG-based program representation cannot allow for a sufficiently granular program representation like that shown in the right half of Fig.~\ref{fig:limitations_of_dag}, this entire for-loop structure must be collapsed to a single DAG node and presented to CEDR as a single unit to be scheduled.
Because it is unlikely that an accelerator exists on the system that can handle this specific sequence of iterated kernels, this single node is likely to only be supported on the CPU, and benefits of acceleration in this application are reduced.

\begin{figure*}[th]
    \centering
    \includegraphics[width=0.75\linewidth]{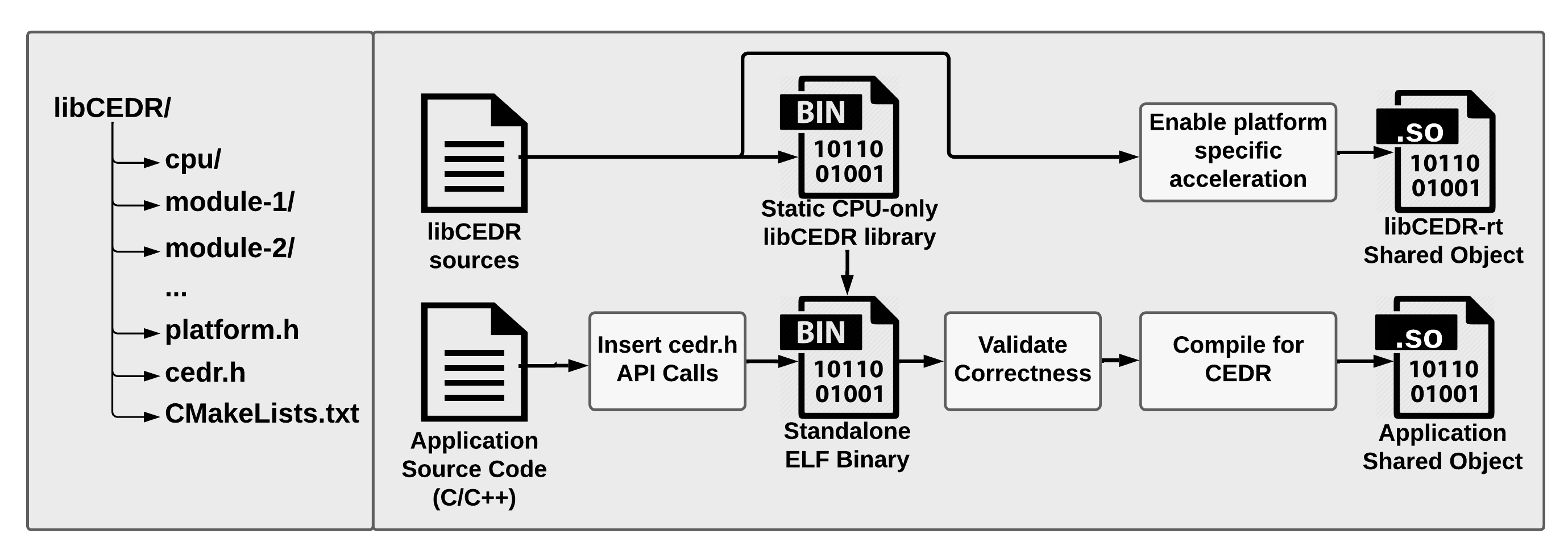}
    \caption{Overview of the libCEDR and CEDR-API compilation methodology}
    \label{fig:api_compilation_overview}
    \vspace{-4mm}
\end{figure*}

\subsection{Contributions of this Work}

To address these limitations, in this work, we adjust CEDR's frontend tooling to consist of a new API-based development workflow shown in Fig.~\ref{fig:api_compilation_overview}.
One of the core developments is a new \texttt{libCEDR} library and associated \texttt{cedr.h}.
The structure of this library is shown in the left side of Fig.~\ref{fig:api_compilation_overview}.
APIs for use in application code are exposed to developers through the \texttt{cedr.h} header file.
This header contains high level kernel declarations that do not contain any implementation details of the underlying operation with samples provided in Listing~\ref{lst:sample_api_declarations}.
As different DSSoC platforms develop different accelerator implementations of these kernels, they are incorporated through \emph{libCEDR Modules}.
As an example, for a platform with a Fast Fourier Transform (FFT) accelerator, \texttt{libCEDR} provides an \texttt{fft} module.
This module is then responsible for providing physical implementations of the high-level APIs as desired.
Finally, the \texttt{platform.h} header file provides global information about the platform in use such as base addresses for accelerators' AXI4~\cite{AXI4} interfaces to enable driverless memory-mapped I/O (MMIO) control and dispatch of tasks.
It is expected that all APIs in this library provide, at a minimum, standard C/C++ implementations that can be leveraged across all the platforms that CEDR supports.
As such, at compilation time, the user configures \texttt{platform.h}, chooses which \emph{libCEDR Modules} to enable, and receives two outputs: a static \texttt{libcedr.a} archive containing only the CPU C/C++ implementations of all implemented APIs and a ``runtime'' \texttt{libcedr-rt.so} shared object containing both the CPU C/C++ implementations and all chosen accelerator implementations through their respective \emph{libCEDR Modules}.

\begin{lstlisting}[
    firstnumber=1,
    basicstyle=\scriptsize\ttfamily,
    label=lst:sample_api_declarations,
    caption=Sample \texttt{libCEDR} API Declarations,
    float=tbh,
    escapechar=\%
]
  void CEDR_FFT(data_t* input, data_t* output, 
                size_t size, bool forward_transform);
  void CEDR_GEMM(data_t* A, data_t* B, data_t* C,
                size_t ROW_A, size_t COL_A, size_t COL_B);
  void CEDR_CONV2D(data_t* input, size_t height, 
                size_t width, data_t* mask, 
                size_t mask_dim, data_t* output);
\end{lstlisting}

Once \texttt{libCEDR} is compiled, a user moves through the workflow shown in the right side of Fig.~\ref{fig:api_compilation_overview}.
One key benefit of this compilation approach is the way that it enables rapid application bring up and evaluation prior to testing in complex heterogeneous computing scenarios.
In early stages of development integration, a user can begin by treating \texttt{libCEDR} like any other CPU-based library.
Once they are convinced that their application is functionally correct, they can compile for CEDR-API by simply building as a shared object that avoids linking in implementations for their \texttt{libCEDR} API calls.
This shared object application is then provided to the runtime using the same IPC submission flow shown in Fig.~\ref{fig:cedr_runtime}.

During startup, the CEDR Runtime is provided with the corresponding \texttt{libCEDR-rt} shared object containing all of the system's API implementations, and it builds a mapping from each API and resource type pairing to a physical implementation of that API on that resource if one exists.
When an application is received by the CEDR Runtime's main event loop, the shared object is parsed and a new system thread is spawned that executes that application's \texttt{main} function.
As this main function executes, it periodically encounters \texttt{libCEDR} API calls.
These calls are linked during binary parsing against implementations in \texttt{libCEDR-rt} that themselves call an \texttt{enqueue\textunderscore kernel} function inside the CEDR runtime.
When \texttt{enqueue\textunderscore kernel} is invoked, a task is placed into CEDR's ready queue, and from there, all of CEDR's existing heuristics are able to process it in the same fashion as the DAG-based methodology.
As this process is multi-threaded in nature -- involving the user application's thread, the main CEDR thread, and the eventual worker thread of the executing resource -- there is a need for a synchronization methodology that the user application can rely on to acknowledge completion of the underlying API call.

\begin{figure}[tbh]
    \centering
    \includegraphics[width=0.8\linewidth]{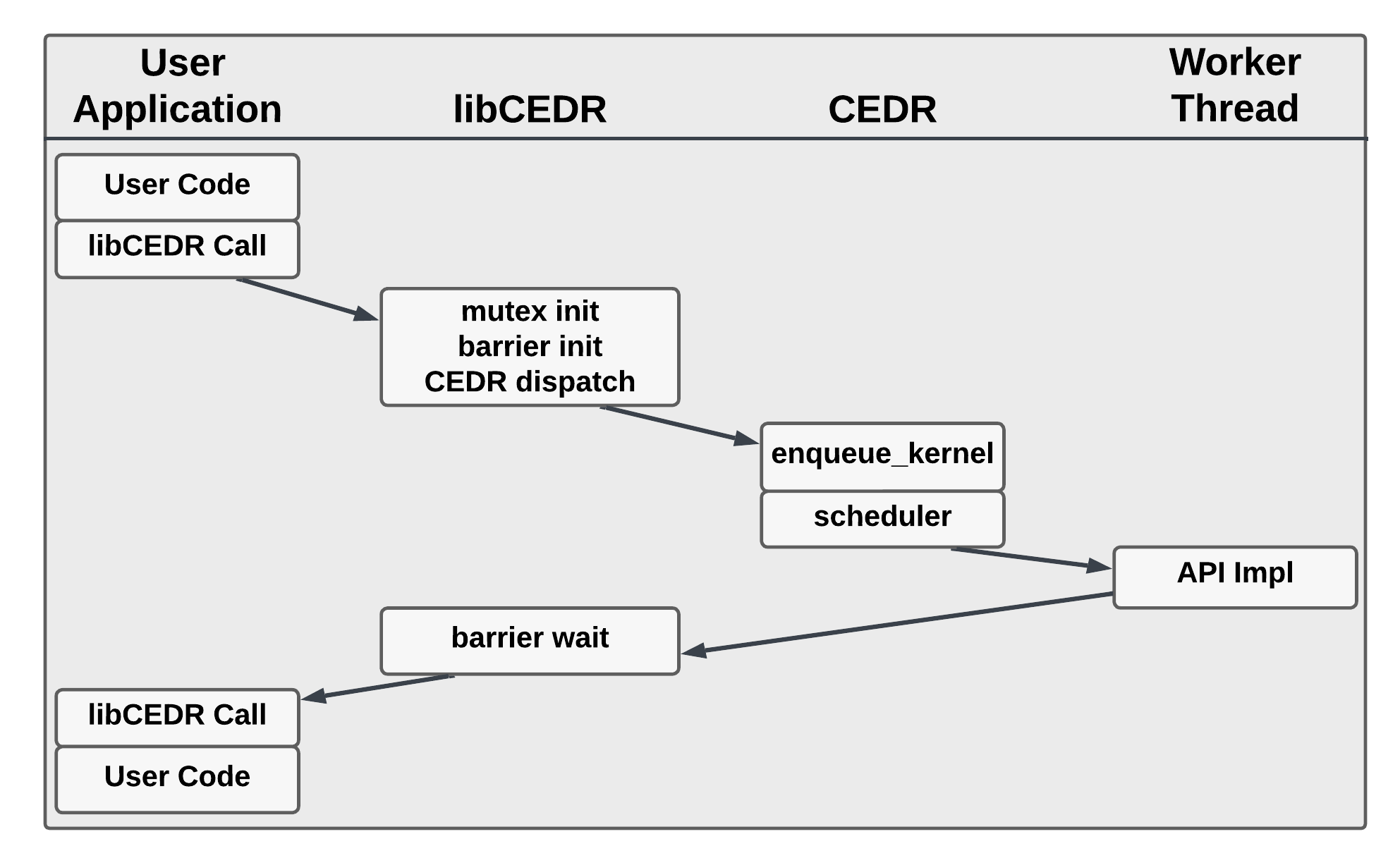}
    \caption{Synchronization methodology dispatching heterogeneous kernels from \texttt{libCEDR} to CEDR}
    \label{fig:synchronization}
     \vspace{-2mm}
\end{figure}

An overview of this synchronization methodology is shown in Fig.~\ref{fig:synchronization}.
Before pushing the requested task to the ready queue, the user's application thread initializes a set of \texttt{pthread\textunderscore cond} and \texttt{pthread\textunderscore mutex} variables to use to receive updates on the progress of its task.
After dispatching the new task with CEDR, the thread then reaches a \texttt{pthread\_cond\_wait} barrier and goes to sleep.
As the task propagates through to the scheduler and eventual API implementation, the corresponding worker thread signals task completion back to the application thread via \texttt{pthread\_cond\_signal}.
After this, the application thread wakes and resumes its computation.

While this methodology ensures that we preserve functional correctness relative to the baseline application's single-threaded execution, it does leave performance to be desired for high-performance users as it does not allow them to parallelize execution of kernels.
To support these needs, we have also developed non-blocking variations of all APIs that are present in \texttt{libCEDR}.
These APIs allow the end user to have full control over the task synchronization primitives such that they can manually maximize parallelism in the underlying application.
It has been designed such that it is still compatible with the standalone validation to shared object compilation workflow shown in Fig.~\ref{fig:api_compilation_overview}.
As we will see in Section~\ref{sec:overhead_analysis}, these non-blocking APIs allow users to extract equivalent performance to the DAG-based methodology without sacrificing productivity improvements and support for richer program representations.

\section{Experimental Setup} \label{sec:experimental-setup}
For our experiments, we use the Xilinx Zynq Ultrascale+ ZCU102~\cite{ZCU102} and NVIDIA Jetson AGX Xavier~\cite{Xavier} development boards.
We use three representative real-world applications for our evaluations covering radar processing, communications system, and autonomous vehicle domains with Pulse Doppler (PD), WiFi TX (TX), and Lane Detection (LD). 
Pulse Doppler calculates velocity of an object, by measuring distance of the object using 256-point FFTs, and measuring the frequency shift between transmitted and emitted signals. 
WiFi TX generates packets of 64 bits and prepares for transmission over an arbitrary channel through scrambler, encoder, modulation, and forward error correction processes. WiFi TX relies on 128-point inverse FFT for each packet transmitted.
Lane Detection is a convolution intensive routine from autonomous vehicles domain. 
In the literature, it has been shown that implementing convolution in the frequency domain rather than the spatial domain through a combination of FFT and pointwise product (ZIP) operations can reduce algorithmic complexity and inference time~\cite{abtahi2018accelerating}.

A workload composed of these three applications allow us to evaluate various scenarios where a heterogeneous SoC is shared by multiple applications in an interleaved manner. An example scenario could involve Lane Detection running as a continuous process where Pulse Doppler and WiFi TX applications arrive dynamically and are executed periodically.
Such scenarios allow us to study the relationship between degree of SoC heterogeneity, scheduling overhead and quality of schedules achieved by various heuristics targeting autonomous vehicles domain. 
The Lane Detection application stresses the FFT accelerator on the emulated heterogeneous SoC with number of 1024-point FFTs and IFFTs scaling to 16384 and 8192  instances respectively for a 960x540 image.
WiFi TX and Pulse Doppler are lower latency applications with number of FFTs scaling to 100 and 512 respectively.
Driven by these three applications, we use FFT and ZIP as key functions that are supported with accelerator based execution. Each application is implemented via the hardware agnostic API calls for each function.

We compile all three applications using the CEDR compilation toolchain described in Section~\ref{sec:cedr-api} and prepare binaries to be executed on heterogeneous SoC configurations that are emulated on both ZCU102 and Jetson development boards. The ZCU102 is formed of 4 ARM cores running at 1.2 GHz and programmable FPGA fabric where we invoke FFT accelerators running at 300 MHz. The FFT accelerator is implemented using Xilinx FFT IP supporting up to 2048-point FFTs. The FFT accelerator uses direct memory access (DMA) to manage data transfers between ARM cores and accelerator through AXI4-Stream~\cite{AXI4}. The Jetson board is formed of 8 ARM cores running at 2.3 GHz and a Volta GPU running at 1.3 GHz, where we implement FFT and ZIP accelerators as CUDA kernels. The data transfers between ARM cores and the accelerators are handled with standard \texttt{cudaMemcpy} functions using the PCIe interface. 

We compose heterogeneous SoCs by varying the number and types of processing elements on the ZCU102 platform from the pool of 3 ARM cores along with 8 FFT accelerators. We utilize the Jetson platform to demonstrate the portability of our compile and runtime system and conduct cross-platform comparisons in terms of factors that contribute to the runtime and scheduling overhead. 
The amount of data processed by an application is considered a frame, measured in Megabits (Mb). Injection rate is defined as the rate at which frame instances are generated per second and measured in Mbps.
We use 29 injection rates between 10 and 2000 Mbps, where each injection rate defines a periodic rate of job along with its associated input data arrival for the given workload. 
We use Round Robin (RR), Earliest Finish Time (EFT), Earliest Task First (ETF), and the runtime variant of the Heterogeneous Earliest Finish Time (HEFT\textsubscript{RT}~\cite{mack_performant_2022-nonblind}) scheduling heuristics executed along with the CEDR management thread using one of the ARM cores on the target SoC.

We use metrics of average scheduling overhead per application and average execution time per application for performance evaluation. The scheduling overhead captures the time spent by the runtime in making scheduling decisions. This time is proportional to the number of scheduling rounds made by the runtime as well as the complexity of the scheduling algorithm.
The application execution time is the time difference between the beginning and completion of an application's execution, including the overhead of all scheduling decisions in between. Lower execution times indicate the scheduler’s capability to manage the workload efficiently. 
To make these two metrics comparable across different runtime configurations, we normalize them with respect to the number of applications and take their average over 25 trials to reduce the effect of noise. 
For brevity, throughout the article, we will take each metric (for instance ``execution time'') to refer to its corresponding averaged-per-application version (``average execution time/application'').

\section{Experiments} \label{sec:experiments}
\subsection{Runtime and Scheduling Overhead Analysis}
\label{sec:overhead_analysis}
\begin{figure}
    \centering
    \includegraphics[width=0.9\linewidth]{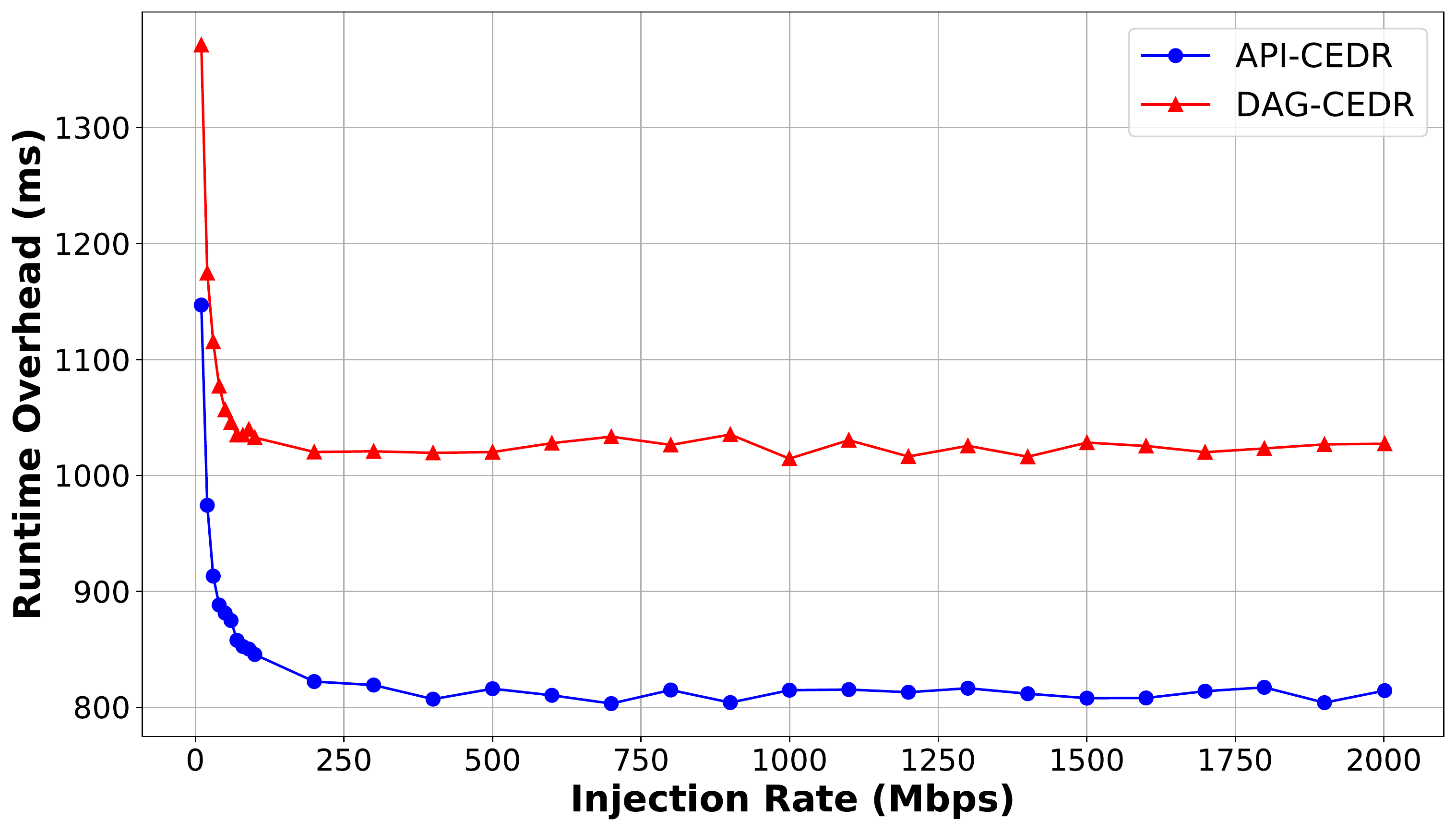}
    \caption{Runtime overhead in API and DAG-based CEDR.}
    \label{fig:runtime_overhead}
     \vspace{-6mm}
\end{figure}

We present the runtime overhead of API-based CEDR with respect to the DAG-based CEDR as illustrated in Fig.~\ref{fig:runtime_overhead}. We define the runtime overhead as the overall time spent by CEDR to receive, manage, and terminate all applications in a given workload. This overhead excludes the overhead of task scheduling, as we define it separately as scheduling overhead.
For this experiment, we use 5 instances for each of the Pulse Doppler and WiFi TX applications as a workload and collect runtime overhead across the sweeping range of the injection rate on the ZCU102 platform with 3 ARM CPUs and 1 FFT accelerator. The X-axis shows the injection rates, and Y-axis shows the runtime overhead. 
As the injection rate grows, the runtime overhead reduces and then saturates at around injection rate of 200 Mbps for both API and DAG-based CEDR. 
The applications arrive in an increasingly overlapping manner to the runtime with the increase in injection rate and, in turn, ready queue size grows. This gives runtime the opportunity to manage multiple tasks concurrently rather than serially, which in turn enables the runtime to complete processing same number of applications in a shorter span of time, thereby reducing the runtime overhead.
The saturation of the trend lines indicate that beyond a certain injection rate, the runtime becomes oversubscribed, where all the applications within the workload get executed by CEDR with maximum concurrency.
We further observe from this plot that, throughout the saturated region, the API-based CEDR achieves a 19.52\% reduction on average in runtime overhead with respect to the DAG-based CEDR. This reduction can be attributed to the simplification of runtime steps in API-based CEDR compared to DAG-based CEDR. For the DAG-based CEDR, the runtime overhead involves time required for receiving and parsing application DAG files via IPC to construct application DAG, parsing shared object, pushing tasks to the ready queue, popping completed tasks from the queue, and finally terminating the completed applications. For the API-based CEDR, two factors contribute to the reduced overhead. First, API-based CEDR does not need to parse DAG files when applications are submitted via IPC. Second, pushing tasks to the ready queue is eliminated as it is handled by the application thread.

\begin{figure}[t]
     \centering
     \begin{subfigure}[b]{0.45\textwidth}
         \centering
         \includegraphics[width=\textwidth]{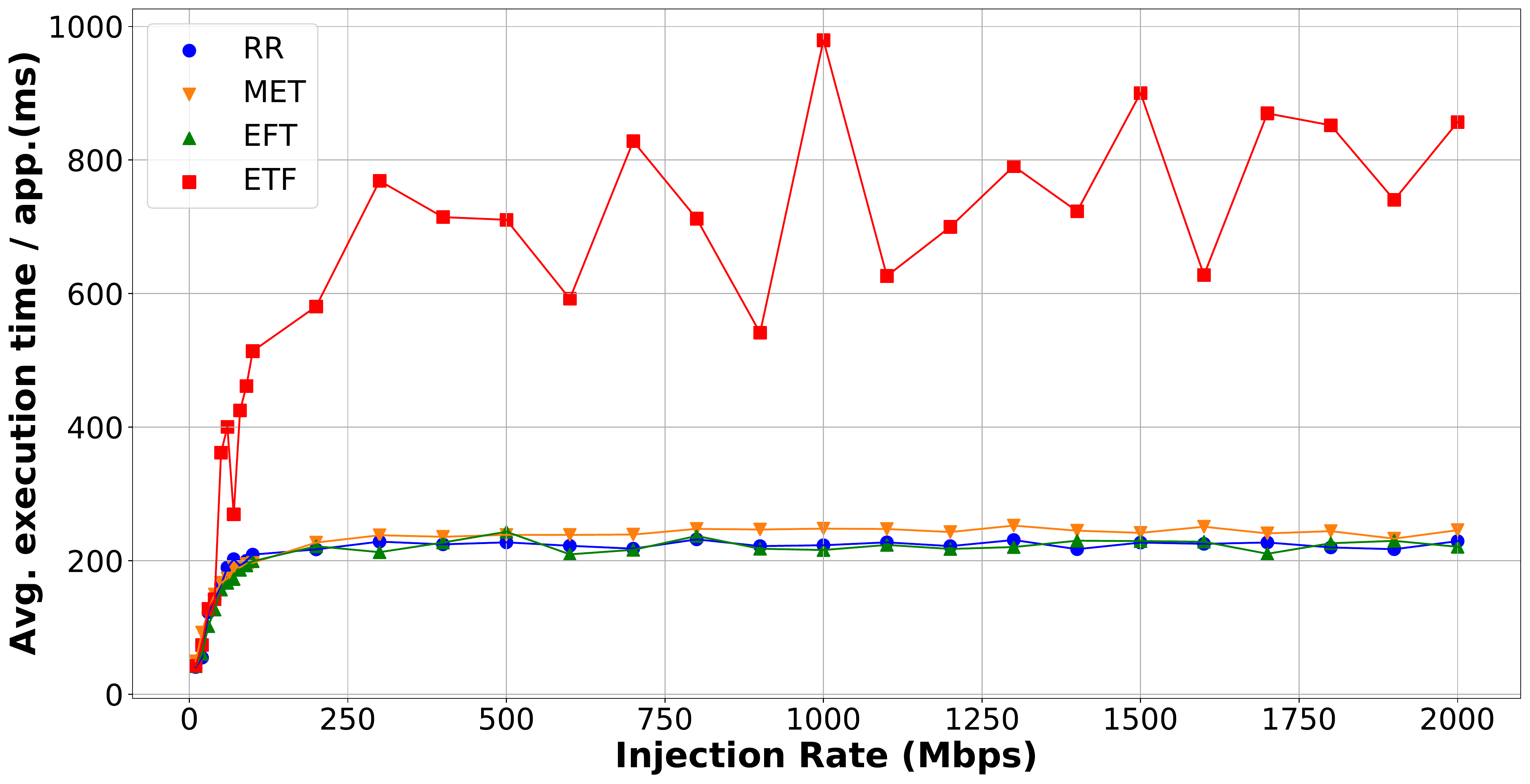}
         \caption{DAG-based CEDR}
     \end{subfigure}
     \hfill
     \begin{subfigure}[b]{0.45\textwidth}
         \centering
         \includegraphics[width=\textwidth]{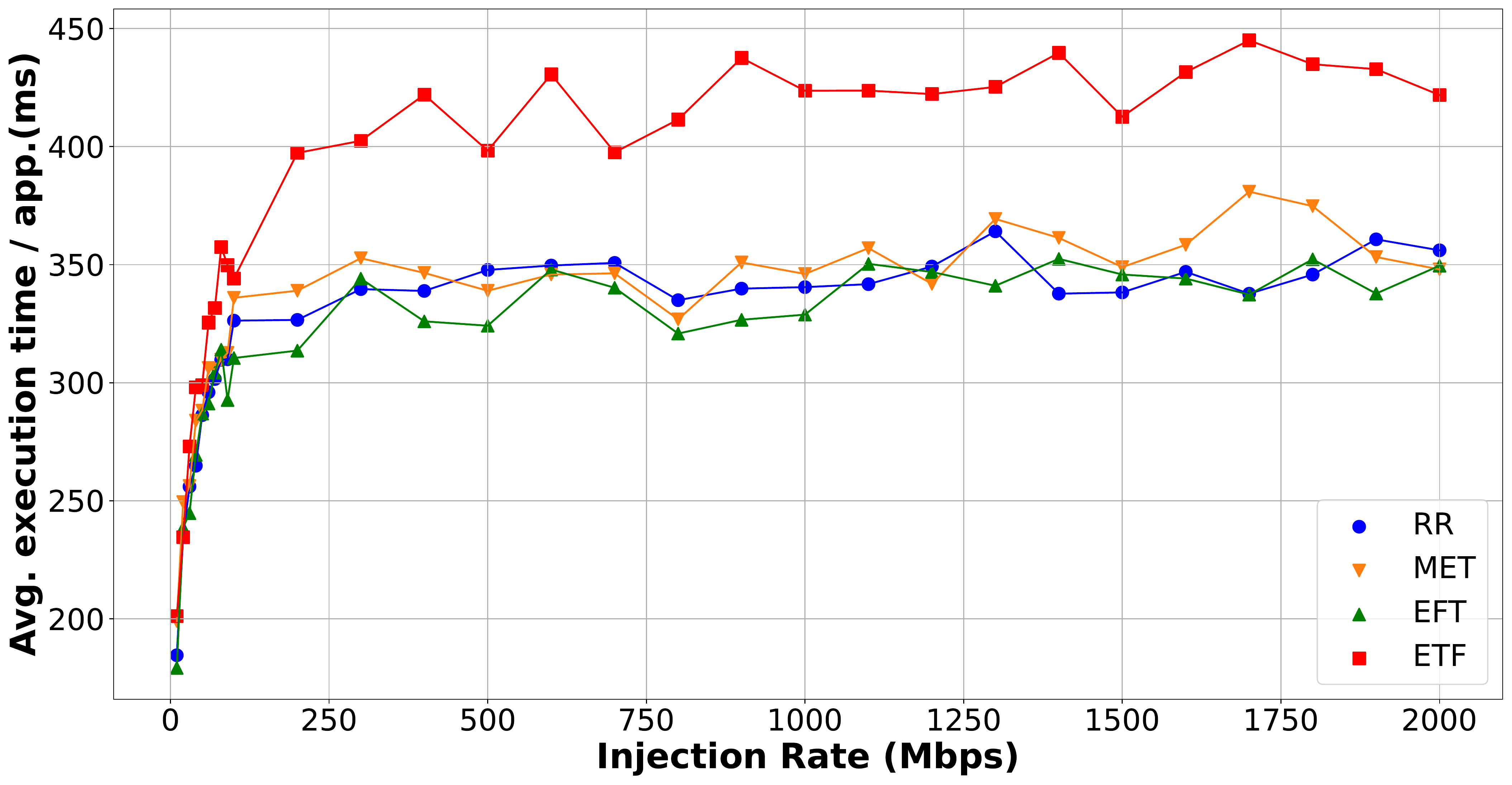}
         \caption{API-based CEDR}
     \end{subfigure}
        \caption{Execution time with respect to injection rate for different schedulers with 3 CPUs, 1 FFT, and 1 MMULT.}
        \label{fig:exec_time_dag_vs_api}
         \vspace{-6mm}
\end{figure}

\begin{figure}[t]
     \centering
     \begin{subfigure}[b]{0.435\textwidth}
         \centering
         \includegraphics[width=\textwidth]{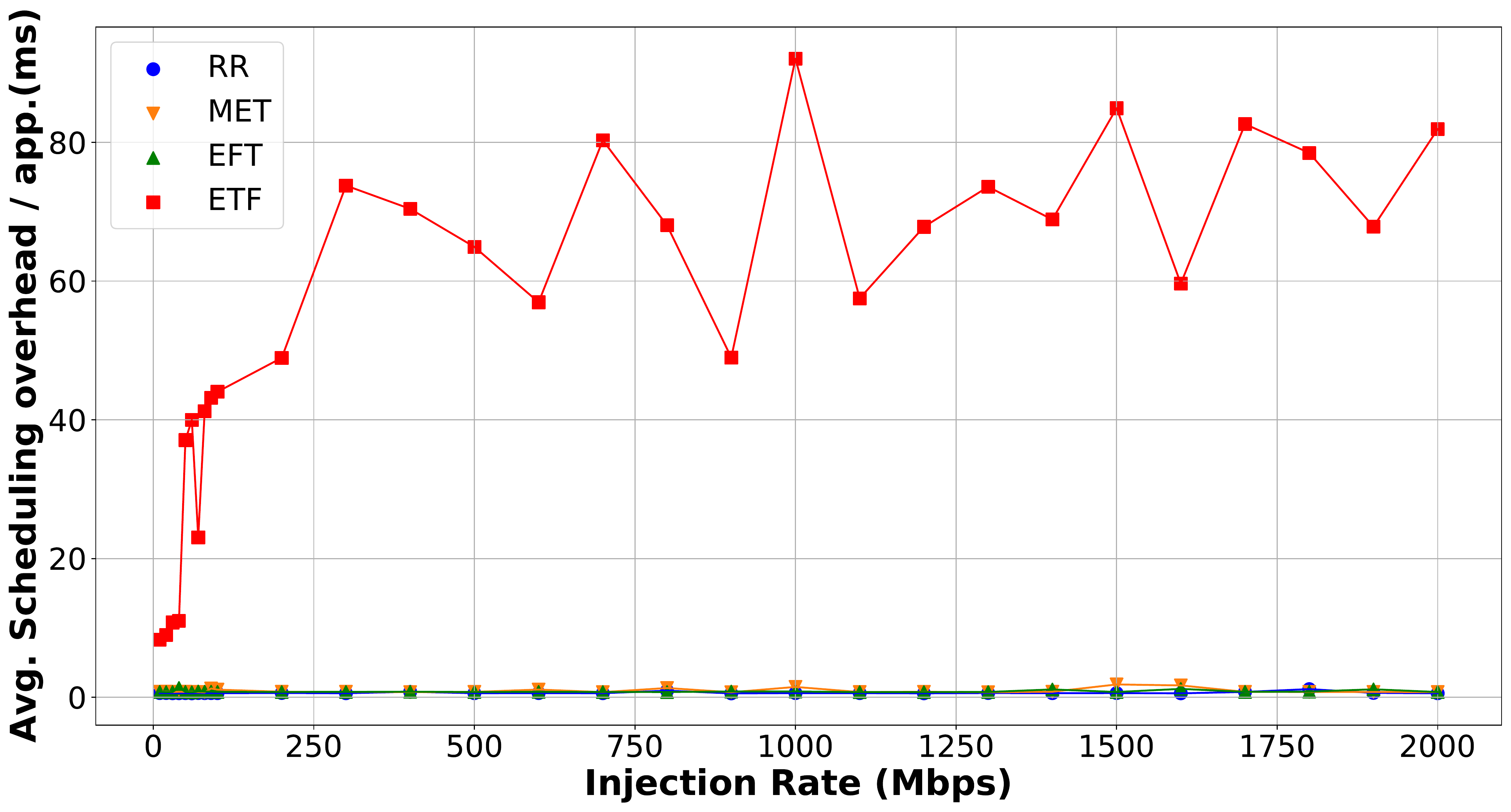}
         \caption{DAG-based CEDR}
     \end{subfigure}
     \hfill
     \begin{subfigure}[b]{0.435\textwidth}
         \centering
         \includegraphics[width=\textwidth]{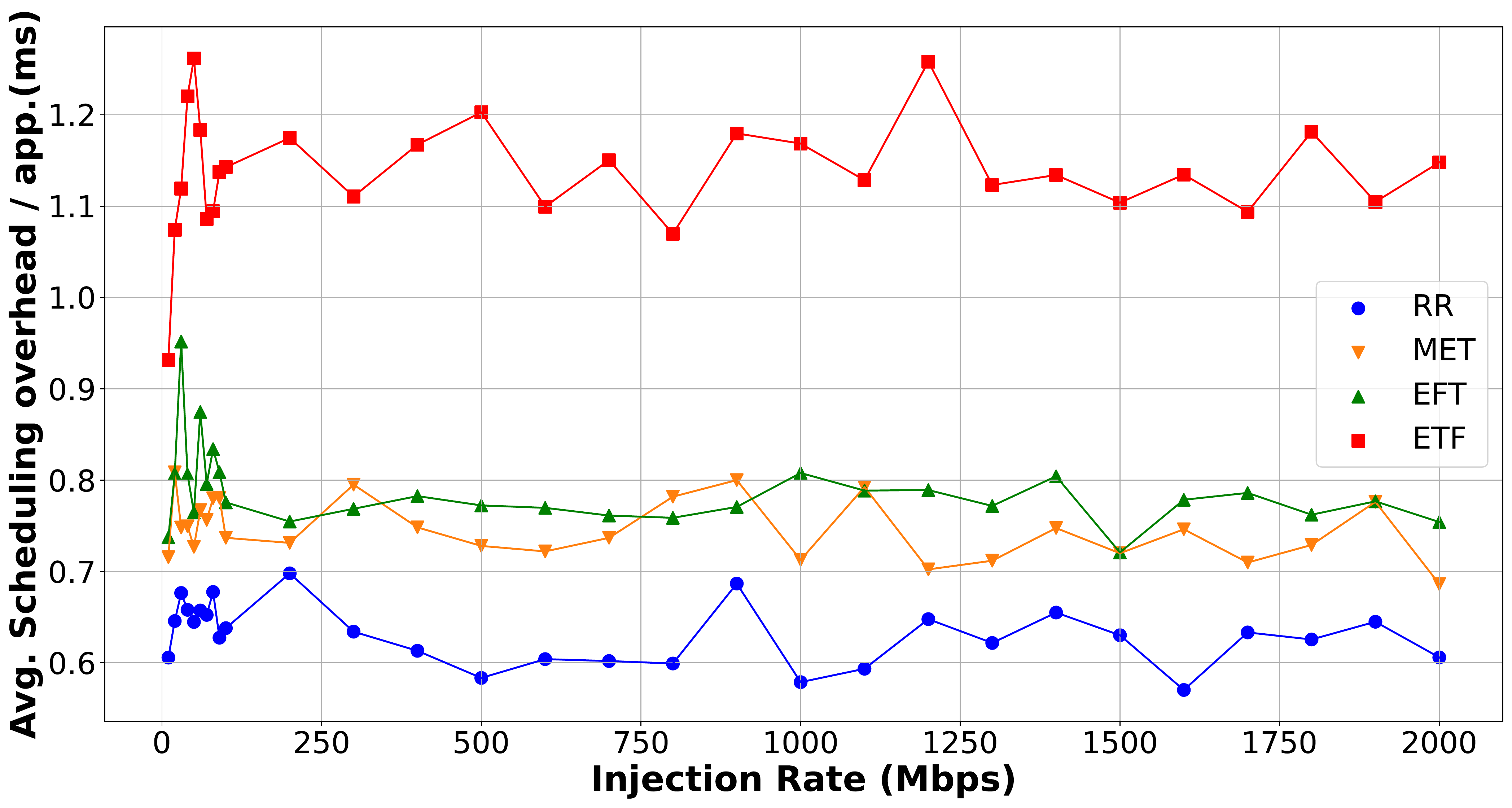}
         \caption{API-based CEDR}
     \end{subfigure}
        \caption{Scheduling overhead with respect to injection rate for different schedulers with 3 CPUs, 1 FFT, and 1 MMULT.}
        \label{fig:sched_time_dag_vs_api}
         \vspace{-6mm}
\end{figure}

Next we present an experiment aiming to validate the application execution time and scheduling overhead trends in API-based CEDR against DAG-based performance trends presented in~\cite{mack_cedrtecs_2023-nonblind}. 
For this experiment, we borrow key parameters such as hardware configuration, workload composition, and scheduling heuristics from the DAG-based CEDR work, for fairness of comparison. The hardware is composed of 3 ARM CPUs, 1 FFT, and 1 MMULT accelerators on the ZCU102 platform. 
We use the same workload as utilized in~\cite{mack_cedrtecs_2023-nonblind} that consists of WiFi TX and Pulse Doppler applications with 5 instances each following the experimental procedure described in Section~\ref{sec:experimental-setup}.
Fig.~\ref{fig:exec_time_dag_vs_api} (a) and (b) show average execution time per application for the DAG and API-based execution on CEDR respectively, with respect to injection rate, where individual line plots represent execution using different schedulers. 
Both Figures~\ref{fig:exec_time_dag_vs_api} (a) and (b) show similar saturation trends as injection rate increases where system becomes oversubscribed at around 200 Mbps. Furthermore, the mean magnitude of the saturated region of API-based execution deviates by 32\% compared to the one of DAG-based execution. 
From the scheduler perspective, we notice ETF scheduler resulting in a significantly higher execution time in both plots while remaining schedulers perform similar to each other in each setup. 
We note that the ETF scheduler in the oversubscribed region shows average execution time as 700 ms in the DAG-based CEDR, while we observe 425 ms with the API-based CEDR. We attribute this execution time reduction for ETF to the ready queue size being smaller, as API-based CEDR only schedules \texttt{libCEDR} API calls/portions of the application, which have support for heterogeneous execution. In DAG-based CEDR, the whole application, including non-accelerated regions, is divided into tasks that are scheduled by CEDR scheduler.

We show the scheduling overhead with respect to injection rate and different schedulers in Figures~\ref{fig:sched_time_dag_vs_api} (a) and (b) for DAG and API-based CEDR executions respectively. 
In both plots we notice that with the exception of the ETF scheduler, the scheduling overhead is stable for the remaining schedulers across the injection rates with very close overhead values. The ETF scheduler however shows remarkably different trend in the API-based CEDR, where the scheduling overhead reduces to around 1.15 ms in the saturated region, from a scale of around 70 ms in DAG-based CEDR. This reduction is due to fewer number of tasks that need scheduling in the API-based CEDR. This further demonstrates that the ETF's execution overhead is more sensitive to the ready queue size than the remaining schedulers. 

Referring back to Fig.~\ref{fig:exec_time_dag_vs_api} (a) and (b), while ETF is observing a reduction in average execution time with the API-based CEDR, the remaining schedulers observe an increase in execution time from around 200 ms on the DAG-based CEDR to around 350 ms on the API-based CEDR in the oversubscribed region. This is primarily due to the way the worker and application threads are managed in API-based CEDR compared to DAG-based CEDR. In DAG-based CEDR, the whole application code is executed on the worker threads as DAG task nodes, hence the available CPU cores are only shared among worker threads. In API-CEDR however, both application and worker threads are launched on the available CPU resources, where only the worker threads execute the application portion with heterogeneity support. 
For the presented experiment on ZCU102 with 3 CPU cores, DAG-based CEDR spawns 4 worker threads while API-based CEDR launches an additional 10 application threads (5 instances of each application), leading to increased thread contention on the underlying CPUs.

We perform the same experiment on the Jetson with a configuration of 3 CPU cores and 1 GPU as shown in Fig.~\ref{fig:exec_time_dag_vs_api_jetson}. 
With the availability of a total of 7 CPU cores, the 4 worker threads (3 CPU and 1 GPU) and 10 application threads have more resources to share between them. This reduces the thread contention compared to the ZCU102. Compared to DAG-based CEDR in Fig.~\ref{fig:exec_time_dag_vs_api_jetson}(a) which spawns only 4 worker threads to execute the workload while under utilizing the available CPU cores, API-based CEDR better exploits the available resources through concurrent execution of worker and application threads, resulting in reduced execution time in Fig.~\ref{fig:exec_time_dag_vs_api_jetson} (b).
We further study the thread contention in Section~\ref{sec:scaling} to better understand the root cause of this behavior.

\begin{figure}[t]
     \centering
     \begin{subfigure}[b]{0.425\textwidth}
         \centering
         \includegraphics[width=\textwidth]{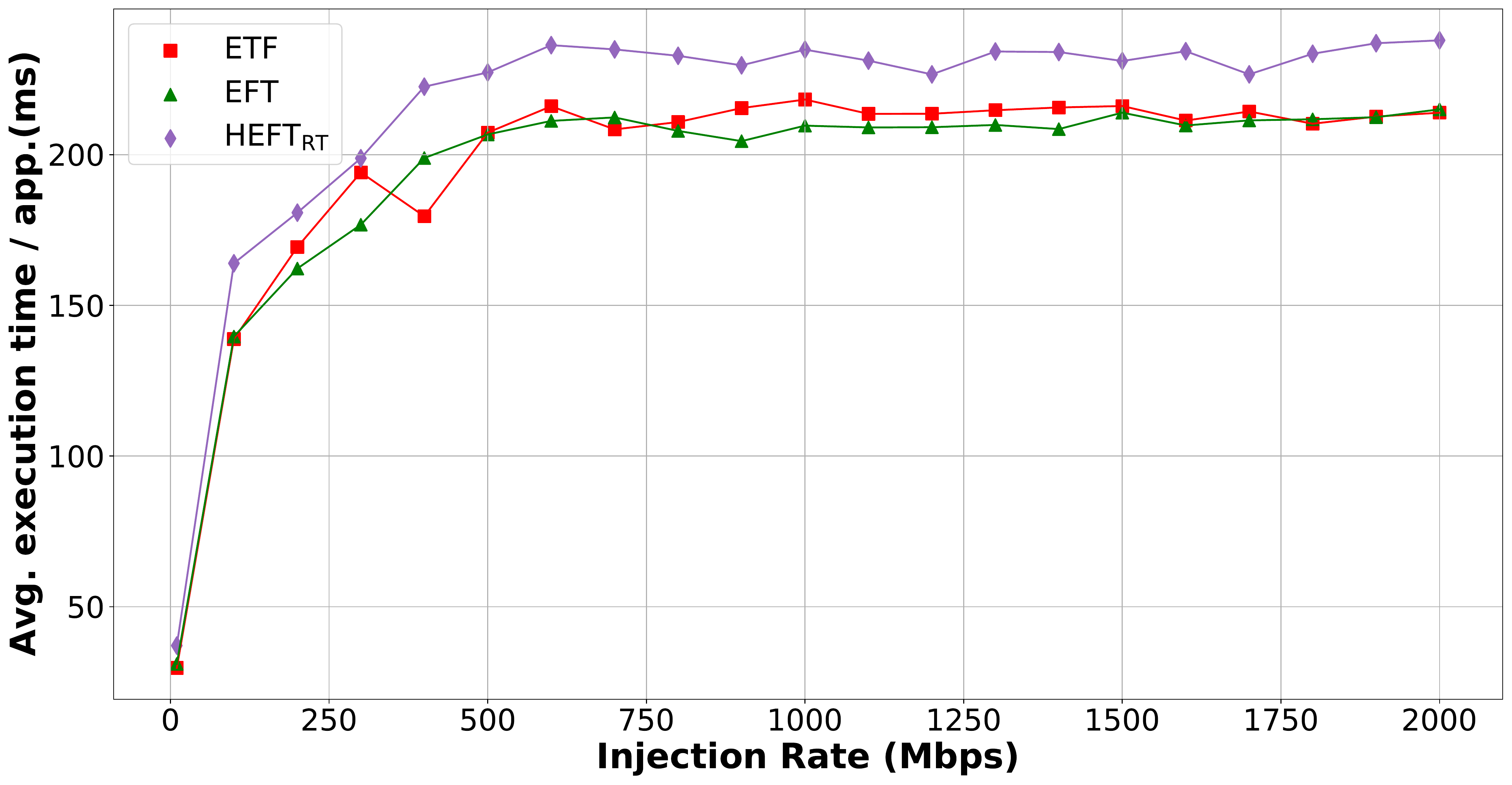}
         \caption{DAG-based CEDR}
     \end{subfigure}
     \hfill
     \begin{subfigure}[b]{0.425\textwidth}
         \centering
         \includegraphics[width=\textwidth]{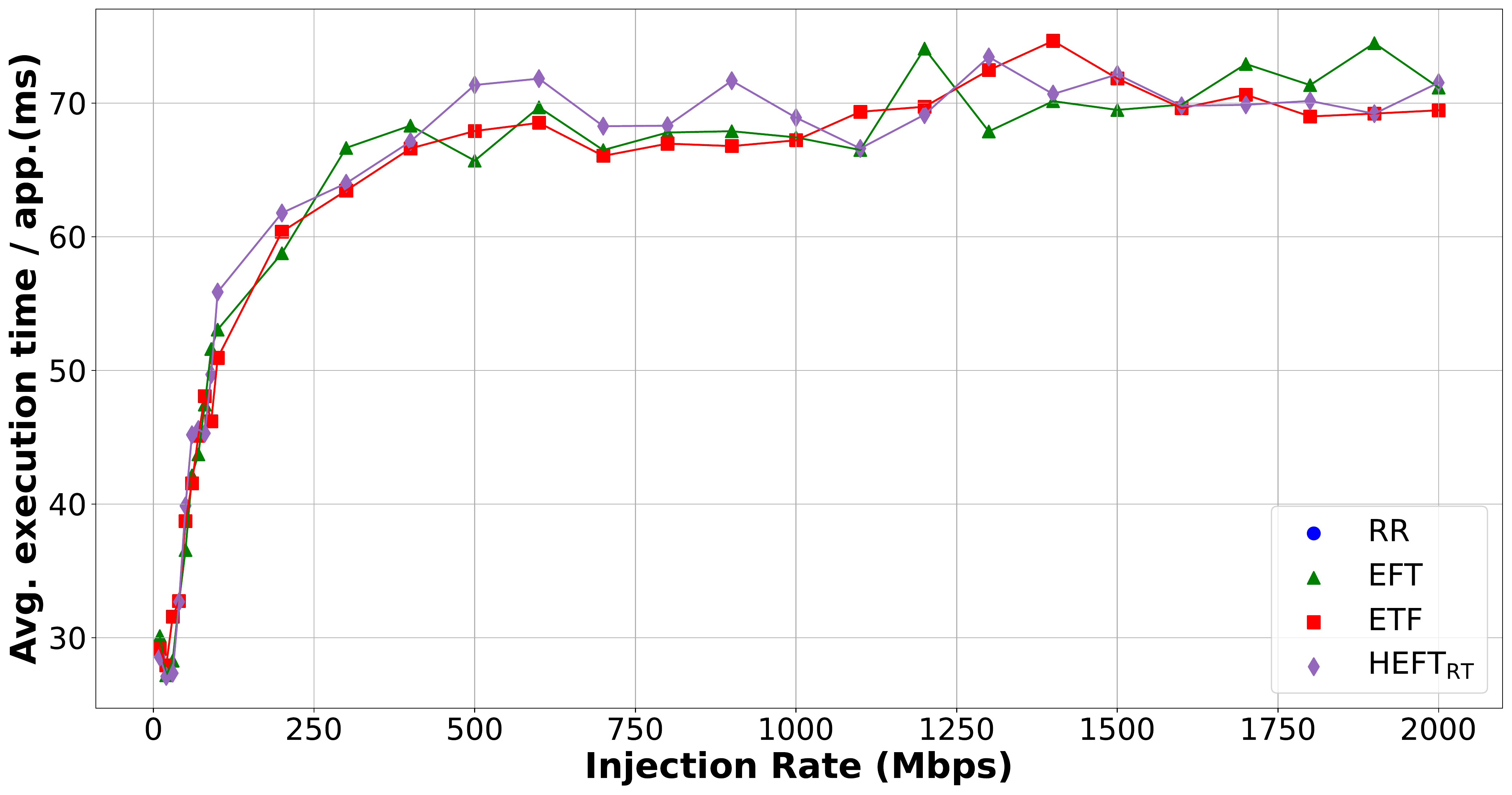}
         \caption{API-based CEDR}
     \end{subfigure}
        \caption{Execution time over injection rates for different schedulers on Jetson AGX Xavier with 3 CPUs and 1 GPU.}
        \label{fig:exec_time_dag_vs_api_jetson}
         \vspace{-6mm}
\end{figure}

For the case of the ETF scheduler on the ZCU102, we notice execution time reduction, because as shown in Fig.~\ref{fig:sched_time_dag_vs_api} (a) and (b), ETF is most sensitive to the heterogeneity with the highest scheduling overhead.
The benefit of reduced ready queue size due to API-based execution results with reduction in scheduling time that is larger in magnitude than the increase in execution time due to thread contention.

\begin{figure}[t]
    \centering
    \begin{subfigure}[b]{0.425\textwidth}
        \centering
        \includegraphics[width=\textwidth]{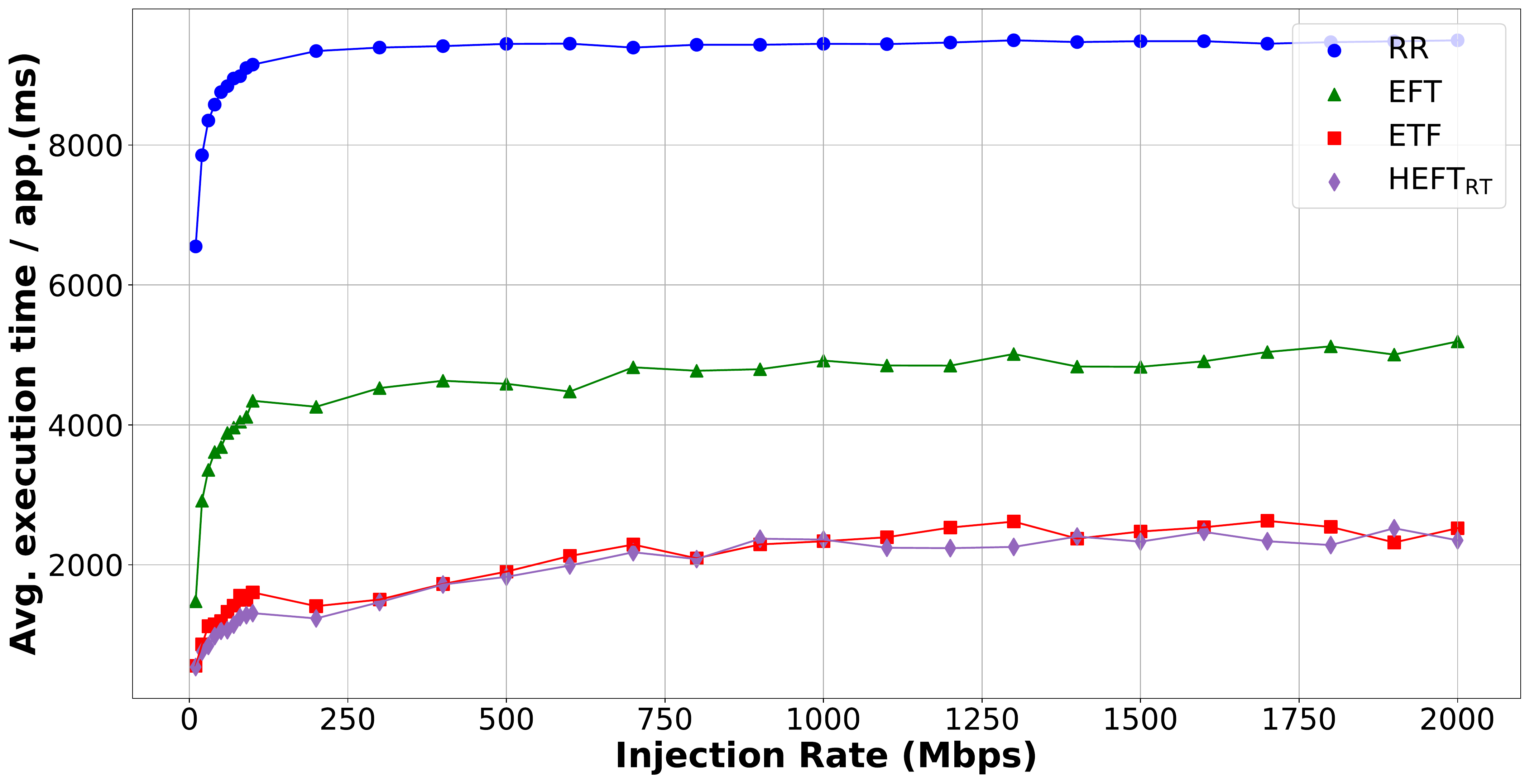}
        \caption{ZCU102: 3 CPU, 8 FFT}
    \end{subfigure}
    \begin{subfigure}[b]{0.425\textwidth}
        \centering
        \includegraphics[width=\textwidth]{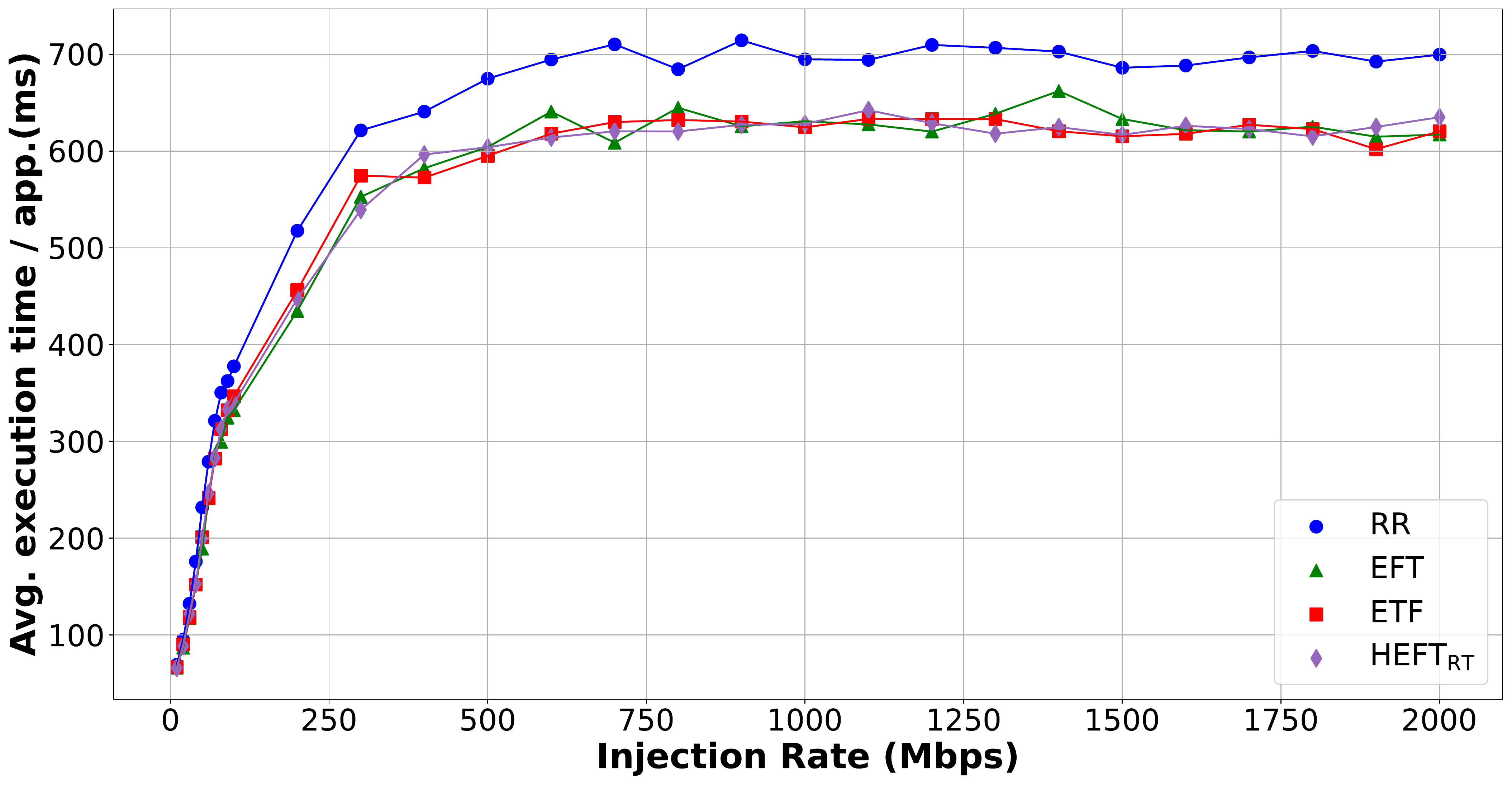}
        \caption{Jetson: 7 CPU, 1 GPU}
    \end{subfigure}
    \caption{Execution time with respect to injection rate in API-CEDR using Lane Detection, Pulse Doppler and WiFi TX.}
    \label{fig:versatility}
     \vspace{-6mm}
\end{figure}

\subsection{Versatility}
\label{sec:versatility}
In this section we expand our experimental evaluations to demonstrate the versatility of the CEDR framework by introducing Lane Detection as a new application to the workload, increasing the number of FFT accelerators on the ZCU102 to 8, and performing execution time performance analysis with respect to changes in injection rate using the same workload on the Jetson platform. As discussed in Section~\ref{sec:experimental-setup}, Lane Detection has a large number of FFT instances, and we expect it will stress both the runtime system and the schedulers as the ready queue size is expected to grow substantially.
The autonomous vehicle workload includes a single instance of Lane Detection as a long latency job while lower latency WiFi TX and Pulse Doppler applications arrive dynamically. Fig.~\ref{fig:versatility} (a) and (b) present the execution time trends of this workload with respect to injection rate, on the ZCU102 and Jetson platforms respectively.

In both execution scenarios, across all schedulers we notice similar saturation trends as observed in Fig.~\ref{fig:exec_time_dag_vs_api}. However, even though number of accelerators has increased from one to eight FFTs, the runtime now approaches the saturation point earlier at 100 Mbps on the ZCU102 based emulation indicating the increase in the complexity of the workload with the inclusion of the LD that stresses both the runtime system and the schedulers. 
In Fig.~\ref{fig:versatility}(b), we observe that for all schedulers the execution time performance shows saturation after 500 Mbps, showing the Jetson platform's ability to cope with the workload better than the ZCU102. While in the best case, during the saturated region ZCU102 platform achieves execution time of 2000 ms, the Jetson based execution is in the range of 600 to 700 ms. 

In both Fig.~\ref{fig:versatility} (a) and (b), RR performs worse than the other schedulers. The performance of RR degrades relative to its performance shown in Fig.~\ref{fig:exec_time_dag_vs_api} (b) where the hardware composition is 3 CPUs and 1 FFT only. RR being a fair scheduler is not able to exploit the increased resource pool or heterogeneity. Whereas the sophisticated and heterogeneity aware schedulers ETF and HEFT\textsubscript{RT} have more opportunity to make better decisions, resulting in lower execution time.

The comparative scheduler performance on the ZCU102 and Jetson platforms presented in Fig.~\ref{fig:versatility} (a) and (b) show different trends.
The performance gap between schedulers on the Jetson is much smaller than the gap we observe on the ZCU102. 
Similar to the conclusion presented in Section~\ref{sec:overhead_analysis}, we believe this is due to the Jetson platform offering richer set of CPU cores that makes it less prone to the thread contention. In the following section we will perform scalability analysis to expose this issue.

\subsection{Scalability}
\label{sec:scaling}

\begin{figure}[t]
     \centering
     \begin{subfigure}[b]{0.45\textwidth}
         \centering
         \includegraphics[width=\textwidth]{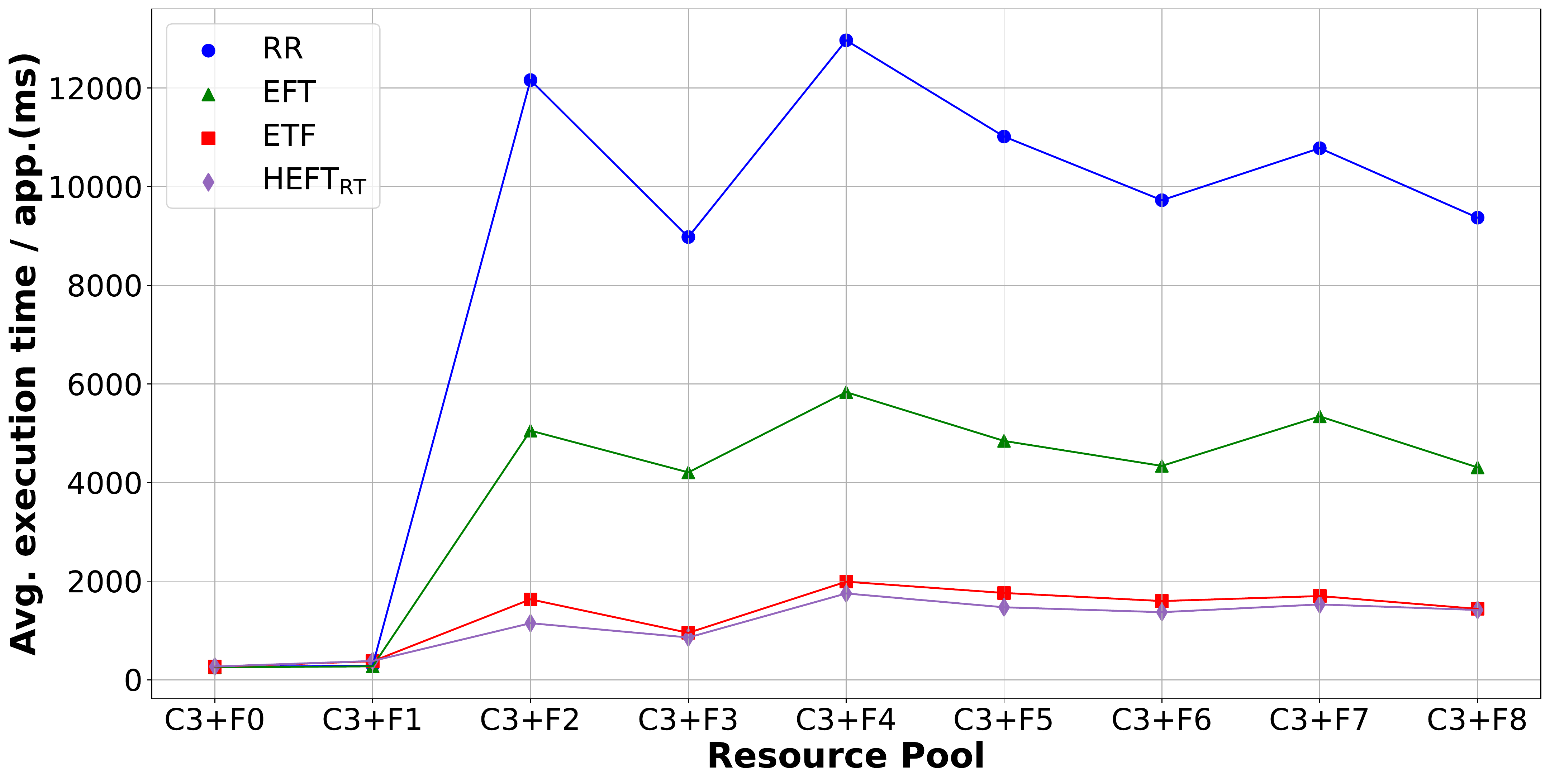}
         \caption{ZCU102: 3 CPU, 0-8 FFT; 300 Mbps injection rate}
     \end{subfigure}
     \hfill
     \begin{subfigure}[b]{0.45\textwidth}
         \centering
         \includegraphics[width=\textwidth]{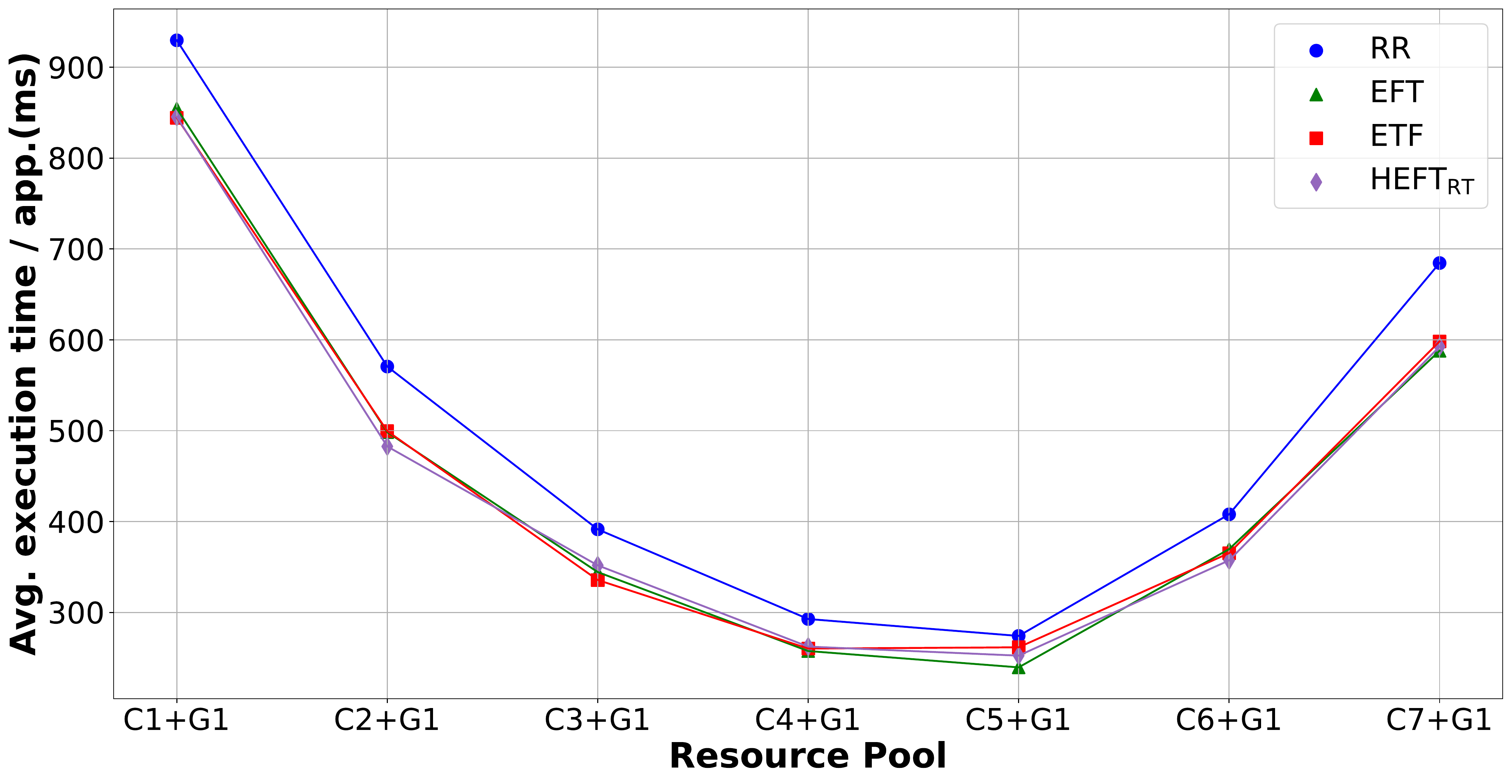}
         \caption{Jetson 1-7 CPU, 1 GPU; 500 Mbps injection rate}
     \end{subfigure}
        \caption{Average execution time per application with respect to varying PE configurations.}
         \vspace{-6mm}
        \label{fig:scalability_exec}
\end{figure}

In this section, we study the impact of increasing the number of processing elements on the application execution performance in CEDR. 
Fig.~\ref{fig:scalability_exec} (a) and (b) show the application execution time of the autonomous vehicle workload on the ZCU102 and Jetson platforms respectively. 
We fix the workload injection rate to 300 Mbps and 500 Mbps when executing applications on the ZCU102 and Jetson platforms respectively, which are the points where these two systems are in oversubscribed state as observed in Section~\ref{sec:versatility}.
The X and Y-axes of both Fig.~\ref{fig:scalability_exec} (a) and (b) present the resource pool and corresponding average execution times respectively.
On the ZCU102 platform in Fig.~\ref{fig:scalability_exec} (a), the resource pool configurations consist of a fixed CPU count of 3 as there are a maximum of 3 ARM cores on the platform, and a varying FFT accelerator count from 0 to 8. For Jetson in Fig.~\ref{fig:scalability_exec} (b), we keep the GPU count fixed at 1 as there is one GPU on the platform, and vary the CPU count between 1 and 7, as there are 7 CPU cores available. One CPU core on both ZCU102 and Jetson is reserved for the execution of CEDR runtime.

The execution scenario on ZCU102 in Fig.~\ref{fig:scalability_exec} (a) shows that the least execution time is observed with 3 CPUs and no FFT accelerators, across all schedulers. Increasing the FFT accelerator count rapidly increases the execution time. This increase can be explained by the sharing of limited CPU cores by larger number of threads. The 3 CPU cores each accommodate the 3 CPU worker threads, the launched application non-kernel threads, as well as the accelerator management threads for the FFTs. 
As illustrated in Fig.~\ref{fig:cedr_runtime}, with the increasing PE count (e.g., FFT), the number of worker threads sharing each CPU core increases, leading to each thread waiting for longer periods to get access to the CPU core. This, in turn, increases the application execution time.
We further notice a gap in the application execution times of different schedulers as FFT count scales up. 
The RR scheduler performs worse than other schedulers, as it tries to use all of the PEs equally. 
This maximizes the number of worker threads competing for the scarce CPU resources, creating contention and in turn results with poor execution time performance. 
The EFT scheduler performs better than RR, as it doesn't force the uniform use of all PEs, rather it focuses on assigning tasks to a subset of PEs that can finish the tasks earliest. The sophisticated heterogeneity aware schedulers ETF and HEFT\textsubscript{RT} further reduce the execution time compared to EFT. ETF not only attempts to find the most optimal task-to-PE mapping during scheduling, but also tries to find the most optimal task to schedule first. This enables ETF to better choose the subset of available PEs that minimize the execution time. 
We find that, in this experiment, HEFT\textsubscript{RT} narrowly achieves the best application execution time performance.

We observe an interesting trend in execution time for all the schedulers on the Jetson platform as we increase the number of CPU cores as shown in Fig.~\ref{fig:scalability_exec} (b). Here, as the CPU worker thread count increases from one to five, the execution time across all schedulers show a downward trend. This reduction is achieved due to the fact that the Jetson has 7 CPU cores available, out of which one is dedicated for GPU management, and remaining 6 CPU cores can accommodate up to 6 CPU worker threads without needing any processor sharing between multiple worker threads. 
Therefore increasing CPU thread count introduces opportunity for concurrent execution and reduces execution time. 
However, this trend is not linear downward, rather it is polynomial with a minimum execution time observed at 5 CPU and 1 GPU threads. This trend is caused by the fact that CEDR-API launches the application non-kernel threads on all 7 CPU cores regardless of the number of worker threads. 
As the number of worker threads increases from 1 CPU to 5 CPUs, more parallel worker threads get introduced, and sharing of the CPU resources between worker and application threads also increases. 
This causes the downward execution time to be polynomial rather than linear. Beyond the 5 CPU 1 GPU configuration, addition of worker threads increases CPU resource sharing between application and worker threads to the point where the overhead of thread sharing diminishes the execution time reduction. Therefore, the execution time experiences polynomial increase.

In our scheduling overhead analysis with respect to increase in number of FFTs on the ZCU102 and number of CPU cores on the Jetson platforms, we observe negligible overhead at a scale of 0.1\% and 0.5\% respectively for each hardware configuration relative to the application execution time. This supports our earlier analysis, that the increase in execution time with increasing resource count is primarily caused by the worker thread contention.

\section{Related Work} \label{sec:related-work}
Many DSSoC design efforts start with simulation-based modeling in both high level simulation~\cite{arda_2020_DS3, matthews_2020_MosaicSim, vega_2020_STOMPTool} and cycle accurate simulation~\cite{cong_2015_PARADECycleaccurate, shao_2016_CodesigningAccelerators, xiao_2019_SelfOptimizing}.
In early design space exploration (DSE) scenarios, many of these options are highly effective at narrowing down the scope of designs that are worth exploring on hardware.
Compared to CEDR, these works are complementary as the designs that are narrowed down via early DSE can then be modeled on commercial FPGA platforms and evaluated to a greater extent with CEDR in order to collect ground-truth hardware measurements that feed back into future cycles of chip design.

Focusing on application runtimes, we can segment the literature by works that target accelerator-rich heterogeneous platforms and those that do not.
Prioritizing discussion of those targeting accelerator-rich heterogeneous platforms, there are many works of note.
Bolchini et al.~\cite{bolchini_opencl_2018} propose a runtime controller for OpenCL-based applications on heterogeneous platforms.
It offers many notable features including the ability to perform cluster-level mapping of tasks on Linux and monitor power or execution metrics.
However, the authors do not discuss the ability to launch simultaneous applications or adjust their scheduling policy.
Picos++~\cite{tan_picos++_2019} proposes a hardware-based runtime that provides support for applications written with OmpSs or OpenMP.
These applications are mapped with Nanos++ API calls~\cite{nanos++} using the Mercurium Source-to-Source compiler.
The current capabilities and status of this compiler is described in~\cite{deharo_2021_OmpSsFPGA}.
While the compilation tooling in this ecosystem is substantial, due to the hardware-based design of their runtime, they are unable to support interchangeable and platform-independent scheduling policies.
Kumar et al.~\cite{kumar_2021_continuum} present DELTA, an approach for dynamically scheduling Function-as-a-Service (FaaS) computations to heterogeneous, network-connected computing resources.
The adaptive, resilient nature of their scheduling heuristics are notable, but distinct from CEDR, their definition of heterogeneity primarily focuses networked collections of CPU-only or CPU-GPU systems.
A large unsolved problem in that area would seem to be development of a FaaS-based methodology for dispatching work to function-specific accelerators like FFTs. 
Hseih et al. have introduced SURF~\cite{hsieh_surf_2019}, a runtime built to allow efficient execution on heterogeneous SoCs.
It follows a similar API-based programming approach to that discussed here, and each API can have a number of implementations on heterogeneous resources.
However, their approach would appear to only support linear chains of kernels and lacks support for parallel dispatch through a non-blocking execution methodology.
Pinto et al. present StarVZ~\cite{pinto_2021_starvz}, a performance evaluation framework built on StarPU~\cite{augonnet2011starpu}, a task-based runtime that supports heterogeneous execution of OpenCL or CUDA-based accelerators.
Vasiliadis, Tsirbas, and Ioannidis~\cite{vasiliadis_2022_manyworlds} propose a device-agnostic, adaptive scheduling approach for scheduling machine learning kernels on heterogeneous architectures.
Kim et al.~\cite{kim_2021_iris} present IRIS, a heterogeneous runtime system that incorporates a unique set of resource discovery, adaptive scheduling, data movement, and programming model capabilities.

While a comprehensive review of this domain is infeasible given page constraints, we believe that CEDR provides a unique set of functionality through its open source accessibility; its status as a highly portable, configurable runtime; and its ability to support rapid DSE for DSSoCs through support for arbitrary workload injection, scheduler integration, PAPI-level performance counters, and kernel-specific accelerators.

\section{Conclusion} \label{sec:conclusion}
In this work, we present CEDR-API, a new programming methodology for the CEDR framework that enables more productive programming of domain-specific architectures while supporting concurrent execution of heterogeneous kernels.
We have expanded CEDR-API to the domain of autonomous vehicles, and we have conducted a number of experiments that explore the overhead, versatility, and scalability characteristics of this modified runtime.
We have found that, while runtime overhead is reduced, as systems evolve into highly heterogeneous architectures with large numbers of distinct processing elements, runtime systems should be designed in ways that allow them to cope with that growth in heterogeneity.

One promising path to address the barrier of CPU availability is to leverage progress in big.LITTLE architectures and exchange a fraction of the heavyweight CPUs with a larger quantity of lightweight CPUs specialized for worker thread management.
In future work, we will explore the development of such optimized, core-rich architectures to enable maximal parallelism across diverse configurations of heterogeneous accelerators while minimizing the energy and latency that such configurations otherwise introduce.

\printbibliography

\end{document}